\documentclass[aps,prc,preprint,amsmath,amssymb,preprintnumbers,superscriptaddress]{revtex4} 

\usepackage{amsmath}
\usepackage{graphicx}
\usepackage{bm}
\usepackage{hyperref}
\usepackage{braket}
\usepackage{multirow}
\usepackage{mathrsfs}
\usepackage{longtable}
\usepackage{pdflscape}

\allowdisplaybreaks[4]

\begin{document}

\title{ Nuclear magnetism in the deformed halo nucleus $^{31}$Ne }

\author{Cong Pan }
\affiliation{State Key Laboratory of Nuclear Physics and Technology, School of Physics, Peking University, Beijing 100871, China}
\affiliation{Department of Physics, Anhui Normal University, Wuhu 241000, China}

\author{Kaiyuan Zhang }
\affiliation{Institute of Nuclear Physics and Chemistry, CAEP, Mianyang, Sichuan 621900, China}

\author{Shuangquan Zhang }
\email{sqzhang@pku.edu.cn}
\affiliation{State Key Laboratory of Nuclear Physics and Technology, School of Physics, Peking University, Beijing 100871, China}

\begin{abstract}

Based on the point-coupling density functional, the time-odd deformed relativistic Hartree-Bogoliubov theory in continuum (TODRHBc) is developed. Then the effects of nuclear magnetism on halo phenomenon are explored by taking the experimentally suggested deformed halo nucleus $^{31}$Ne as an example.
For $^{31}$Ne, nuclear magnetism contributes 0.09 MeV to total binding energy, and  the breaking of Kramers degeneracy results in 0-0.2 MeV splitting in canonical single-particle spectra.
The blocked neutron level has a dominant component of $p$ wave and it is marginally bound. However, if we ignore nuclear magnetism, the level becomes unbound.
This shows a subtle mechanism that nuclear magnetism changes the single-particle energies, causing a nucleus to become bound.
Based on the TODRHBc results, a prolate one-neutron halo is formed around the near-spherical core in $^{31}$Ne. The nucleon current is mostly contributed by the halo rather than the core, except near the center of the nucleus. A layered structure in the neutron current distribution is observed and studied in detail.
\end{abstract}

\date{\today}

\maketitle

\section{Introduction}

The exotic nuclei far away from $\beta$-stability line, exhibiting many phenomena different from stable nuclei, such as halos \cite{Tanihata1985PRL,Minamisono1992PRL,Schwab1995ZPA}, changes of nuclear magic numbers \cite{Ozawa2000PRL} and pygmy resonances \cite{Adrich2005PRL}, have become one of the most fascinating topics in nuclear physics.
Since most exotic nuclei will remain beyond the experimental capability in the foreseeable future \cite{Zhang2019PRL,Ahn2019PRL,Ahn2022PRL}, their properties mainly rely on theoretical predictions.

The relativistic density functional theory has been proved to be a powerful tool in nuclear physics by its successful applications in describing many nuclear phenomena \cite{Meng2016book}, and has attracted wide attention in recent years \cite{Ring1996PPNP,Vretenar2005PR,Meng2006PPNP,Niksic2011PPNP,Meng2013FoP,Meng2015JPG,Zhou2016PS,Shen2019PPNP,Meng2021AAAPS}.
In order to describe exotic nuclei, the effects of pairing correlation and the coupling to continuum should be considered properly \cite{Dobaczewski1984NPA,Dobaczewski1996PRC,Meng1996PRL,Meng1998PRL,Meng1998NPA,Meng2006PPNP,Xia2018ADNDT}.
Since most nuclei in the nuclear chart deviate from spherical shape, the deformation effect also plays an important role.
Based on the relativistic density functional theory, the deformed relativistic Hartree-Bogoliubov theory in continuum (DRHBc) was developed \cite{Zhou2010PRC,Li2012PRC}, with the deformed relativistic Hartree-Bogoliubov equations solved in a Dirac Woods-Saxon basis \cite{Zhou2003PRC}.
With the deformation, pairing correlation, and continuum effects included in a microscopic and self-consistent way, the success of the DRHBc theory has been illustrated in the descriptions of halo structures in B \cite{Yang2021PRL,Sun2021PRC}, C \cite{Sun2018PLB,Sun2020NPA}, Ne \cite{Zhong2022SCP}, Na \cite{Zhang2023PRC_Na} and Mg \cite{Zhang2023PLB,An2024PLB} isotopes, as well as many other applications to a variety of nuclear phenomena  \cite{Li2012CPL,Zhang2019PRC,Pan2019IJMPE,Zhang2020PRC,In2021IJMPE,Sun2021SciB,Zhang2021PRC,Pan2021PRC,He2021CPC,Sun2021PRC_AMP,Zhang2022ADNDT,Choi2022PRC,Kim2022PRC,Sun2022CPC,Sun2022PRC,Pan2022PRC,Zhang2022PRC,Guo2023PRC,Zhang2023PRC_2DCH,Xiao2023PLB,Zhao2023PLB,Mun2023PLB}.

It is interesting to note that in most suggested halo nuclei or candidates from the experiments so far, the neutron or proton number is odd (see Fig.~1 of Ref.~\cite{Zhang2023PRC_Na}).
The odd nucleon leads to the vector current and time-odd field, breaking the time-reversal symmetry.
The time-odd field is often referred to as the nuclear magnetic potential due to its similarity with a magnetic field after nonrelativistic reduction \cite{Yao2006PRC}.
In this case, the Kramers degeneracy no longer holds, making the system more complex to treat.
To avoid such complexity, in many works the equal-filling approximation (EFA) \cite{Perez-Martin2008PRC,Schunck2010PRC} is adopted, where the two configurations of the blocked pair are averaved in a statistical manner, leading to that the corresponding currents cancel each other.
Therefore, under the EFA, the nuclear magnetic potential vanishes, and the system is still time-reversal invariant.
In existing DRHBc studies on odd systems, the EFA is also adopted \cite{Li2012CPL,Pan2022PRC}.

The effects of nuclear magnetism have been well studied for deeply bound or rotational nuclei within the relativistic density functional theory.
Based on the calculations in Refs.~\cite{Rutz1998NPA,Afanasjev2010PRCnonrot,Xu2014NPA}, the binding energy is lowered by about 1 MeV for the odd-mass nuclei near $^{16}$O, and by about 0.1 to 1 MeV for the heavier odd-mass nuclei.
The nuclear magnetism also lowers the odd-even mass staggering by about 10\% to 30\%, further influencing the determination of pairing strength \cite{Rutz1999PLB,Afanasjev2000PRC}.
Nuclear magnetism also plays an important role in nuclear magnetic moments \cite{Yao2006PRC,Hofmann1988PLB,Furnstahl1989PRC,Arima2011Sci,Li2011Sci,Li2011PTP,Li2018FoP} and nuclear rotations \cite{Koepf1989NPA,Koepf1990NPA,Konig1993PRL,Afanasjev2000PRC,Afanasjev2010PRCrot,Zhao2012PRCmag}.
In an exotic nucleus near the drip line, with the nucleon separation energy close to zero and the Fermi energy close to continuum threshold, the nuclear magnetism may change the decay properties and exotic structure \cite{Afanasjev2010PRCnonrot,Kasuya2020PTEP}, and even change the location of drip line.
Therefore, in the fully self-consistent calculation for an odd system, the effect of nuclear magnetism should be considered.
In particular, it would be interesting to study the nuclear magnetism in halo nuclei with a theoretical model combining deformation, pairing correlation, continuum and time-odd effects simultaneously.

To explore the effects of nuclear magnetism on halo phenomenon the nucleus $^{31}$Ne is a good candidate. 
This nucleus is the last bound odd-$N$ isotope of Ne \cite{Lukyanov2002JPG}. 
In 2009, a large Coulomb breakup cross section for $^{31}$Ne was observed, which indicated a soft $E1$ excitation and suggested a halo structure \cite{Nakamura2009PRL}.
Subsequent experiments \cite{Gaudefroy2012PRL,Nakamura2014PRL} supported also the halo in $^{31}$Ne, assigned the spin-parity $3/2^-$, and extracted the one-neutron separation energy of only $0.15_{-0.10}^{+0.16}$ MeV.
The halo in $^{31}$Ne has attracted wide attention on the theoretical side \cite{Horiuchi2010PRC,Urata2011PRC,Sumi2012PRC,Zhang2014PLB,Hong2017PRC,Urata2017PRC,Zhang2022JPG,Takatsu2023PRC}.
Recently, the halo structure in $^{31}$Ne and the reaction cross sections on a carbon target were investigated based on the DRHBc theory \cite{Zhong2022SCP}.

In this work, by incorporating self-consistently time-odd field into the DRHBc theory, the time-odd deformed relativistic Hartree-Bogoliubov theory in continuum (TODRHBc) is developed.
Taking $^{31}$Ne as an example, the halo structure and the effects of nuclear magnetism are explored by studying the neutron and proton single-particle spectra, density and vector current distributions, as well as their decomposition into the parts of the core and valence neutron.
In Section \ref{sec:th}, a brief framework of the TODRHBc theory is introduced. The numerical details are given in Section \ref{sec:num}. The results and discussion are presented in Section \ref{sec:discussion}. Finally, a summary is given in Section \ref{sec:summary}.

\section{Theoretical framework}
\label{sec:th}


In the TODRHBc theory, the nuclear magnetism is self-consistently incorporated into the DRHBc theory \cite{Li2012PRC,Zhang2020PRC,Pan2022PRC}. 
The detailed formalism of the TODRHBc theory will be summarized in a future work, and here a brief introduction is presented.

In the TODRHBc theory, the relativistic Hartree-Bogoliubov equation for nucleon reads \cite{Kucharek1991ZPA}
\begin{equation}
	\begin{pmatrix} \hat{h}_D - \lambda & \hat{\Delta} \\ -\hat{\Delta}^* & -\hat{h}_D^* + \lambda \end{pmatrix}
	\begin{pmatrix} U_k \\ V_k  \end{pmatrix} = E_k \begin{pmatrix} U_k \\ V_k  \end{pmatrix} ,
\end{equation}
where $\lambda$ is the Fermi energy, $E_k$ is the quasiparticle energy, and $U_k$ and $V_k$ are the quasiparticle wavefunctions expanded in a spherical Dirac Woods-Saxon (DWS) basis \cite{Zhou2003PRC,Zhang2022PRC}.
$\hat{h}_D$ is the Dirac Hamiltonian, and in coordinate space,
\begin{equation}
	h_D(\bm{r}) = \bm{\alpha} \cdot (\bm{p} - \bm{V}) + V^0 + \beta(M+S) .
\end{equation}
$S$ and $V^\mu$ are the scalar and vector potentials, respectively, and in the framework of point-coupling density functional,
\begin{align}
	S(\bm{r}) & = \alpha_S \rho_S + \beta_S \rho_S^2 + \gamma_S \rho_S^3 + \delta_S \nabla^2\rho_S , \\
	V^\mu(\bm{r}) & = \alpha_V j^\mu + \gamma_V(j_\nu j^\nu) j^\mu + \delta_V \nabla^2 j^\mu + e\frac{1-\tau_3}{2}A^\mu + \alpha_{TV}\tau_3 j_3^\mu + \delta_{TV}\tau_3\nabla^2j_3^\mu ,
\end{align}
where $(\alpha, \beta, \gamma, \delta)$ are coupling constants, and $\rho_S, j^\mu$ and $j_3^\mu$ are scalar density, vector current and isovector current.
The space-like component of $V^\mu$, i.e., $\bm{V}$, is a time-odd field, which changes sign in a time-reversal transformation and vanishes in even-even nuclei.
$\bm{V}$ is often referred to as the nuclear magnetic potential due to its similarity with a magnetic field after nonrelativistic reduction \cite{Yao2006PRC}.
$\hat{\Delta}$ is the pairing potential
\begin{equation}
	\Delta(\bm{r}_1,\bm{r}_2) = V^{pp}(\bm{r}_1,\bm{r}_2) \kappa(\bm{r}_1,\bm{r}_2),
\end{equation}
where $V^{pp}$ is the pairing force and $\kappa$ is the pairing tensor \cite{Ring1980NMBP}.

For an axially deformed nucleus with spatial reflection symmetry, the nucleon densities and currents as well as potentials are expanded in terms of the Legendre polynomials \cite{Zhou2010PRC,Li2012PRC,Zhang2020PRC,Pan2022PRC},
\begin{equation}
	\label{e:lam}
	f(\bm{r}) = \sum_l f_l(r) P_l(\cos\theta) , \qquad l = 0,2,4,\dots,l_{\max}
\end{equation}
where $l$ is restricted to be only even numbers due to spatial reflection symmetry.

For an odd-$A$ nucleus, the blocking effect of the unpaired nucleon(s) needs to be considered.
As has been explained in detail by Refs.~\cite{Ring1980NMBP,Li2012CPL,Pan2022PRC}, in our work, the blocking effect is realized by the exchange of 
the quasiparticle wavefunctions $(V_{k_b}^*, U_{k_b}^*) \leftrightarrow (U_{k_b}, V_{k_b})$ and that of the energy $E_{k_b} \leftrightarrow -E_{k_b}$, where $k_b$ denotes the blocked quasiparticle state.
Similarly, the blocking effect in a multiquasiparticle configuration can be treated.

\section{Numerical details}
\label{sec:num}

In this work, the TODRHBc calculations are based on the point-coupling density functional PC-PK1 \cite{Zhao2010PRC}, which turns out to be one of the best density functionals for describing nuclear properties \cite{Zhao2012PRCmass,Lu2015PRC,Agbemava2015PRC,Zhang2022ADNDT}.
For the pairing channel, a density-dependent zero-range force with the pairing strength $V_0 = -342.5~\mathrm{MeV~fm}^3$ and a pairing window of 100 MeV is taken. 
The box size $R_{\mathrm{box}} = 20$ fm and the mesh size $\Delta r = 0.1$ fm.
For the Dirac Woods-Saxon basis, the angular momentum cutoff $J_{\max} = 19/2~\hbar$ and the energy cutoff $E_{\mathrm{cut}}^+ = 300$ MeV, and the number of the basis states in the Dirac sea is the same as that in the Fermi sea \cite{Zhou2003PRC}.
The Legendre expansion truncation for potentials and densities in Eq.~\eqref{e:lam} is $l_{\max} = 10$.
The convergence for the above numerical conditions has been checked for the nuclei with $8\leq Z \leq 20$, and the accuracy for total energy is better than $0.01\%$.
In order to calculate an odd-$A$ nucleus, the calculation with each possible orbital blocked is performed independently, and the result with the lowest total energy is taken as the ground state \cite{Pan2022PRC}.

\section{Results and discussion}
\label{sec:discussion}

The suggested halo nucleus $^{31}$Ne \cite{Nakamura2009PRL} is taken as an example to be investigated using the TODRHBc theory with PC-PK1. For comparison, the DRHBc calculation that neglects nuclear magnetism (hereinafter referred to as time-even DRHBc) was also performed with the same numerical details.

It is found the calculated separation energy is negative in the mean-field level for both the TODRHBc ($S_n = -0.37$ MeV) and the time-even DRHBc ($S_n = -0.46$ MeV) calculations. 
The time-odd result is more bound by 0.09 MeV, corresponding to the contribution of nuclear magnetism.
We note that the quadrupole deformation of $^{31}$Ne is found around $0.170$, leading to a considerable rotational energy correction, whereas both calculations predict $^{30}$Ne a spherical nucleus without any rotational correction. For the density functional PC-PK1, the rotational correction was shown to play a significant role in improving its mass description of deformed nuclei~\cite{Zhao2010PRC,Zhang2020PRC,Pan2022PRC}, and therefore should be considered for $^{31}$Ne.
Based on the cranking approximation of the moment of inertia \cite{Zhang2020PRC}, the estimated rotational correction energy of $^{31}$Ne is $0.93$ MeV. After including this correction, the separation energy $S_n$ becomes $0.56$ MeV, slightly larger than the experimental data $0.15_{-0.10}^{+0.16}$ MeV \cite{Nakamura2014PRL} and showing the stability of $^{31}$Ne against one-neutron emission. It is also mentioned that the cranking approximation used here is not suitable for spherical nuclei, while the collective Hamiltonian method is expected to further improve the present results \cite{Sun2022CPC,Zhang2023PRC_2DCH}.

\begin{figure}[htbp]
  \centering
  \includegraphics[width=0.3\linewidth]{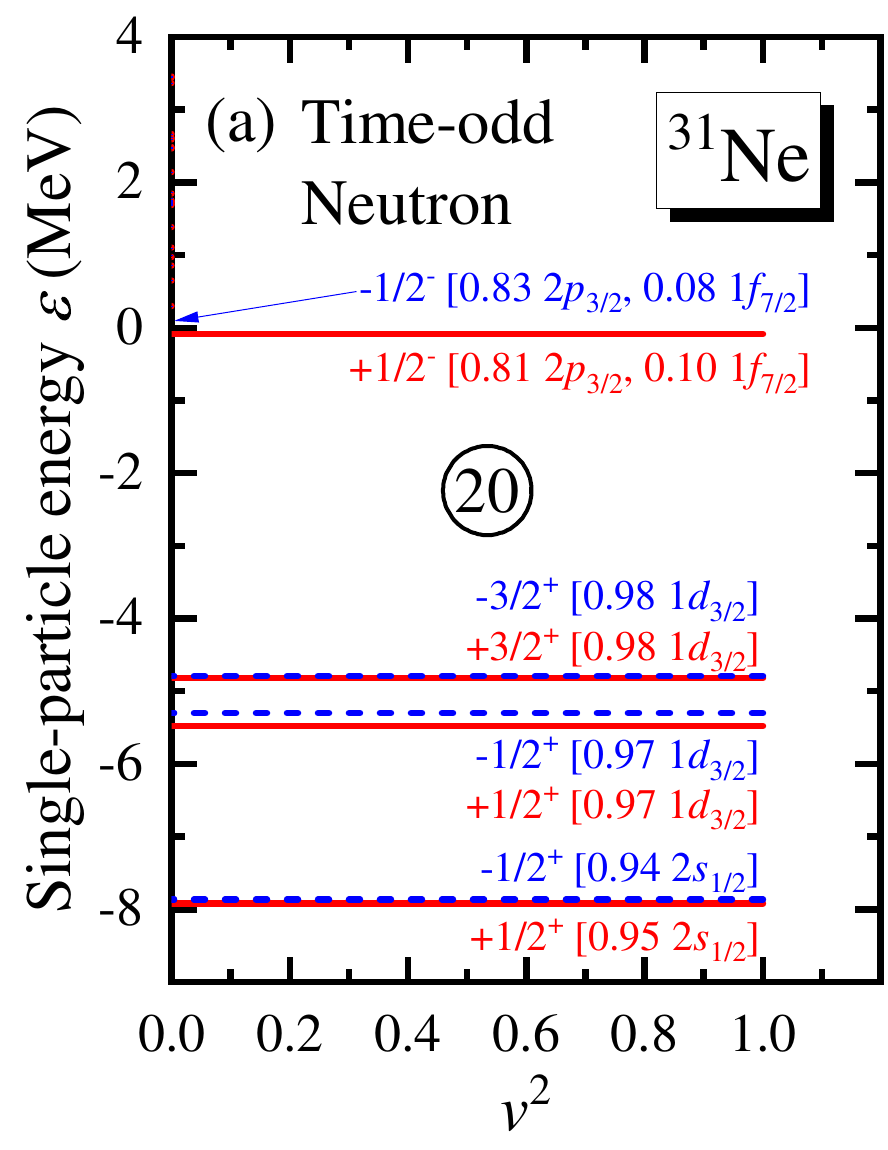}
  \includegraphics[width=0.3\linewidth]{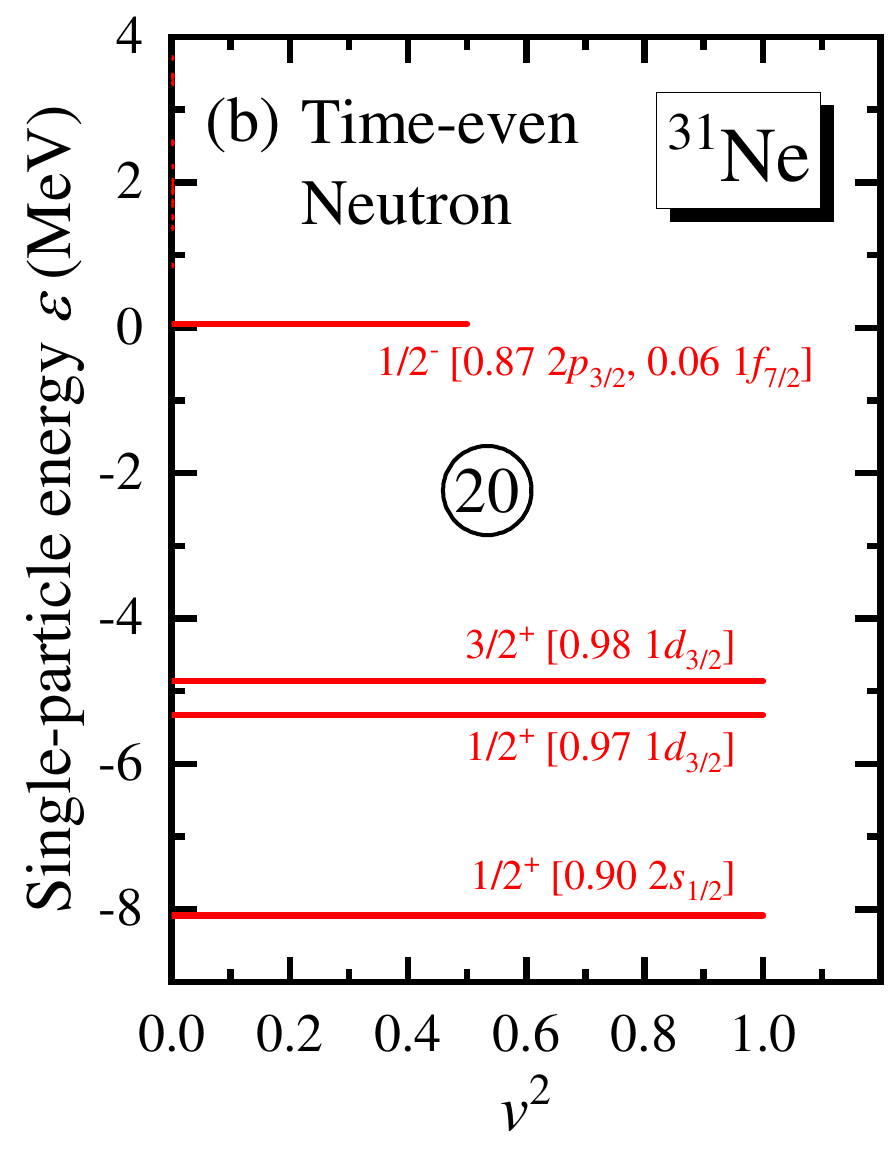} \\
  \includegraphics[width=0.31\linewidth]{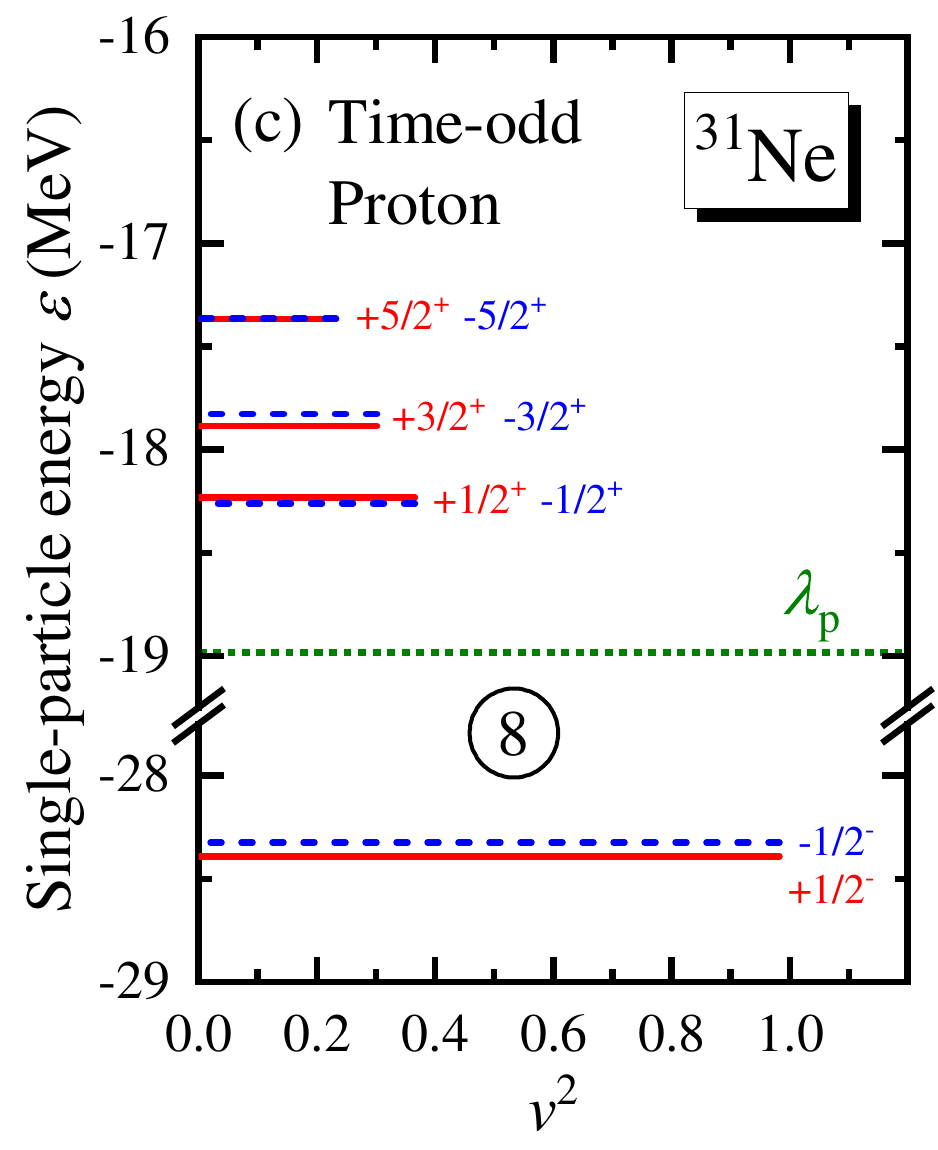}
  \includegraphics[width=0.31\linewidth]{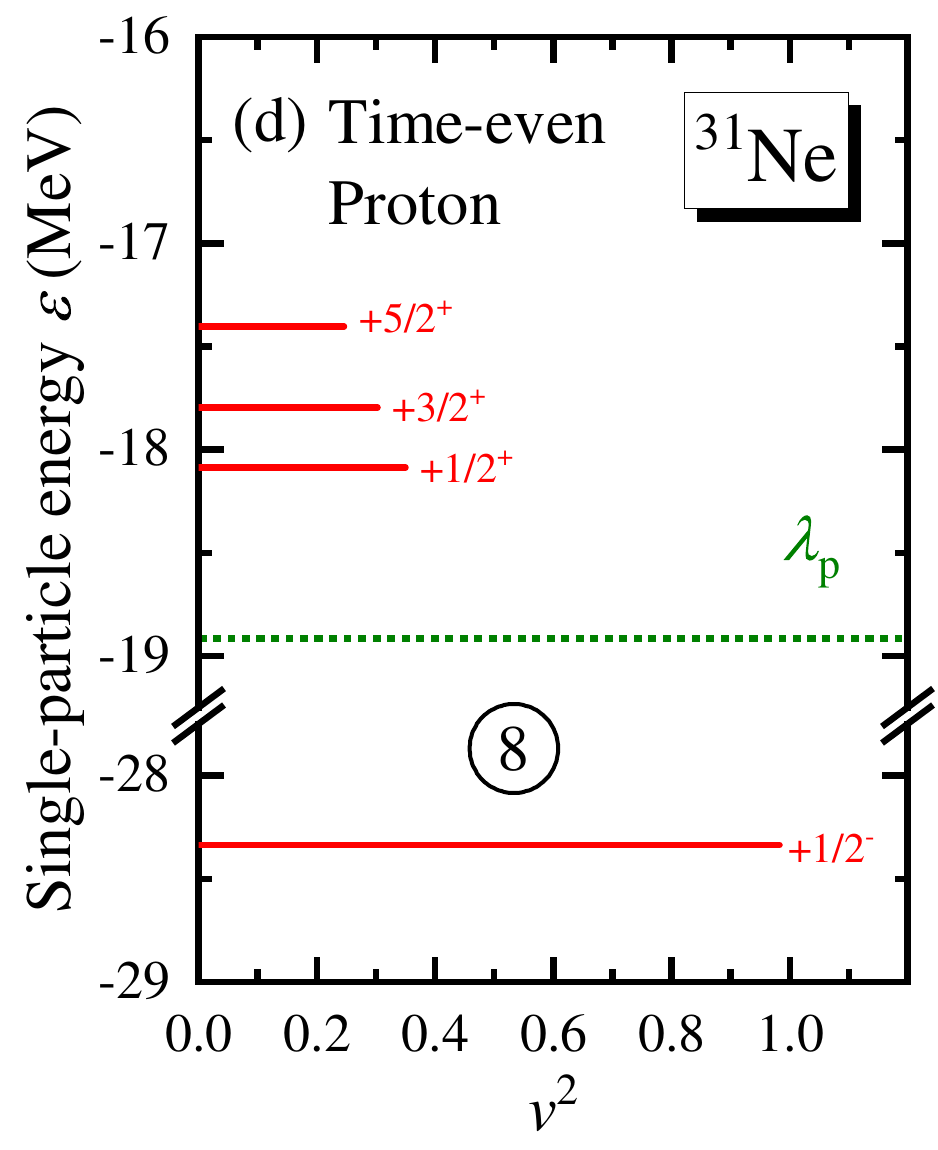}
  \caption{ Single-neutron (a) and proton (c) levels around the Fermi energies in the canonical basis for $^{31}$Ne versus the occupation probability $v^2$ in the TODRHBc calculations with PC-PK1, as well as the corresponding single-neutron (b) and proton (d) levels in the time-even DRHBc calculations.
  Each level is labeled by the quantum numbers $m^\pi$.
  The main components in the DWS basis are given for the neutron levels. The proton Fermi energy $\lambda_p$ is shown with the dotted line. Due to the vanished neutron pairing energy, the neutron Fermi energy equals the energy of the last occupied level, and therefore is not shown here.
   }
  \label{fig_Ne31spl}
\end{figure}

Figure \ref{fig_Ne31spl} show the single-neutron and proton levels around the Fermi energies in the canonical basis for $^{31}$Ne in the TODRHBc calculations with PC-PK1, in comparison with the time-even DRHBc results. In both calculations, the quantum numbers of the blocked neutron orbital are $m^{\pi}=1/2^-$ for the ground state of $^{31}$Ne, and the main components of this valence orbital are $p_{3/2}$, in agreement with the experimental spin and parity $J^\pi = 3/2^-$ \cite{Gaudefroy2012PRL,Nakamura2014PRL}.
As seen, in the time-even DRHBc results, the energies of all neutron and proton levels are doubly degenerate due to the time-reversal symmetry, whereas in the TODRHBc results this degeneracy is broken by the nuclear magnetism with the energy splitting of 0 to 0.2 MeV.
Both calculations yield a vanished neutron pairing energy for $^{31}$Ne. As a result, in TODRHBc, the occupation probability $v^2$ of each neutron level is either 0 or 1, whereas in time-even DRHBc, an exceptional occupation exists for the blocked level with $v^2=0.5$, which stems from the adopted equal-filling approximation \cite{Li2012CPL,Pan2022PRC}. The proton pairing energy doesn't vanish in both calculations.
It is found that although the single-proton energy degeneracy is broken by nuclear magnetism, the $v^2$ of each level is the same with that of its conjugate one.
The proton Fermi energy is decreased by 0.06 MeV after including the nuclear magnetism.

As seen in Fig.~\ref{fig_Ne31spl}, the blocked neutron level in the TODRHBc calculations is the $m^\pi=(+1/2)^-$ one at $\epsilon=-0.087$ MeV, and its dominant component is $p$ wave with above 80\%.
This means in $^{31}$Ne the first neutron level above the major shell $N=20$ is not $f_{7/2}$ dominated, contrary to the order in the traditional spherical shell structure, i.e., $^{31}$Ne resides in the island of inversion \cite{Nakamura2009PRL,Takechi2012PLB}.
Owing to the low centrifugal barrier, a near-threshold level dominated by $s$ or $p$ wave has a relatively larger rms radius and is more diffuse in spatial distribution, which is helpful in the formation of a halo structure and will be discussed further below.
The conjugate state of this blocked neutron level is the unoccupied $(-1/2)^-$ level with a similar component composition, but at a positive energy $\epsilon=0.027$ MeV.
In comparison, the blocked level in the time-even DRHBc calculations is the degenerated $1/2^-$ orbital at $\epsilon=0.049$ MeV, located in continuum.
This indicates that when nuclear magnetism is neglected, $^{31}$Ne is unbound according to the occupation of single-neutron levels.
The inclusion of nuclear magnetism eliminates the degenerate of the $1/2^-$ orbitals, one above and one below continuum threshold, and the occupation of the lower one makes $^{31}$Ne weakly bound.

It should be mentioned that
since the blocked level is very weakly bound, its energy may vary or even cross continuum threshold when the TODRHBc numerical details are changed, even if the numerical convergence has been confirmed with high accuracy.
In addition, Ref.~\cite{Zhong2022SCP} has shown that $^{31}$Ne is bound in the time-even DRHBc theory with other density functionals NL1, NL3 and PK1.
Nevertheless, our results in Fig.~\ref{fig_Ne31spl} demonstrate a possible mechanism of nuclear magnetism that makes an orbital from unbound to bound and a nucleus more stable.


\begin{figure}[htbp]
  \centering
  \includegraphics[width=0.5\linewidth]{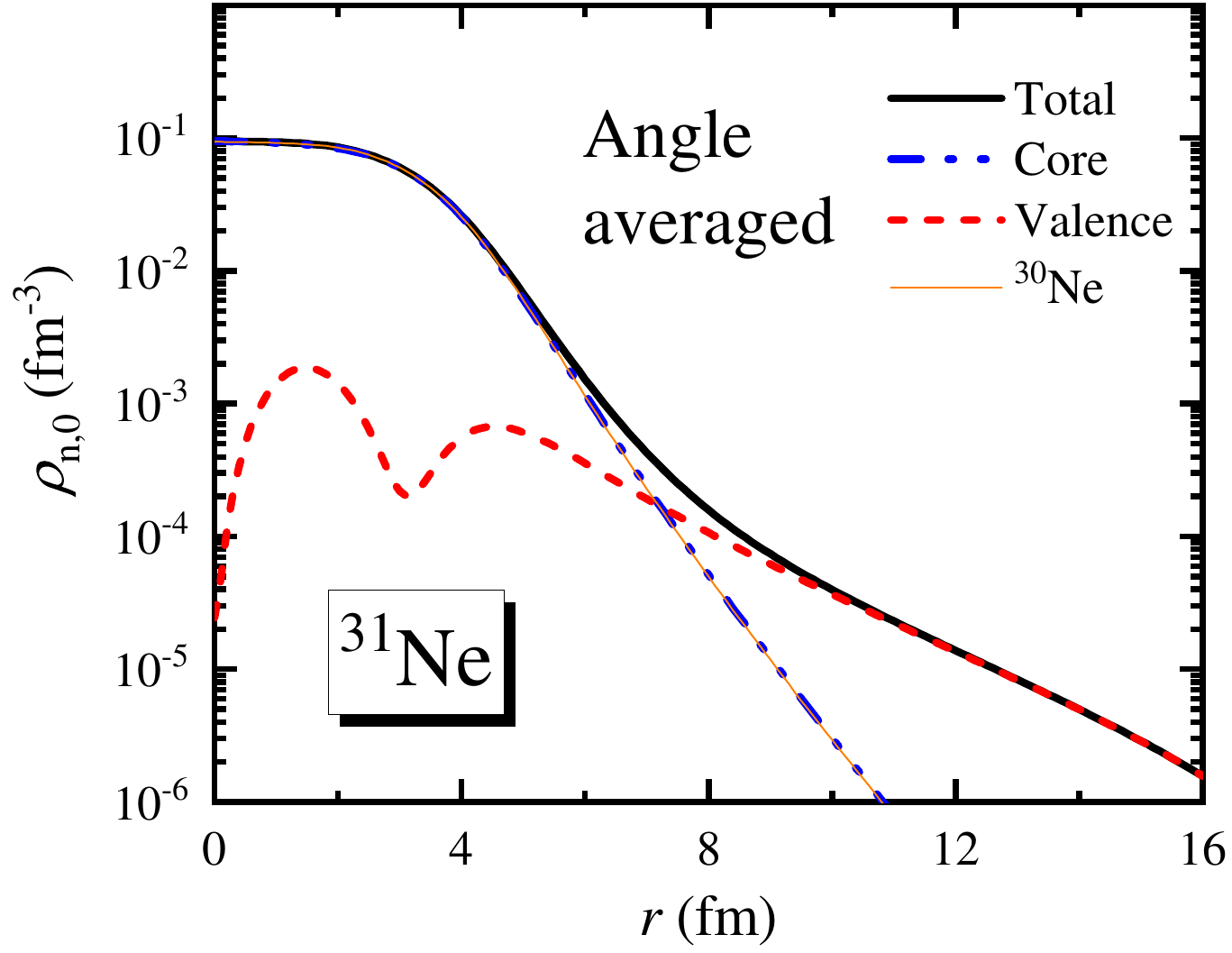}
  \caption{ Angle-averaged neutron density distribution, i.e., the spherical component, for $^{31}$Ne in the TODRHBc calculations, as well as its decomposition into the core and the valence neutron. 
  The density distribution for $^{30}$Ne is shown for comparison.
  }
  \label{fig_Ne31dens1D}
\end{figure}

Figure \ref{fig_Ne31dens1D} shows the angle-averaged neutron density distribution $\rho_{n,0}$ for $^{31}$Ne in the TODRHBc calculations.
According to the single-neutron levels in Fig.~\ref{fig_Ne31spl}(a), there is a pronounced gap larger than 4 MeV between the blocked orbital and the other occupied ones.
Following the strategy in Refs.~\cite{Zhou2010PRC,Li2012PRC,Zhang2020PRC}, $\rho_{n,0}$ of $^{31}$Ne can be decomposed into the core and the valence neutron, as shown by the dashed lines in Fig.~\ref{fig_Ne31dens1D}.
With the increase of $r$, $\rho_{n,0}$ decreases, where the decrease of core part is rapid and that of valence neutron part is relatively slow.
At $r>7.2$ fm, the contribution from valence neutron part becomes the main component of $\rho_n$, forming a long tail with a diffuse density distribution.
Considering the dominant $p$ wave in the valence neutron in Fig.~\ref{fig_Ne31spl}(a), a one-neutron halo is obtained for $^{31}$Ne in TODRHBc calculations.
In Fig.~\ref{fig_Ne31dens1D} the neutron density distribution of the neighboring even-even nucleus $^{30}$Ne is also given, and is found to be similar to the core part of $^{31}$Ne, indicating that $^{31}$Ne is a ``$^{30}$Ne$+1n$'' system.


\begin{figure}[htbp]
  \centering
  \includegraphics[width=0.6\linewidth]{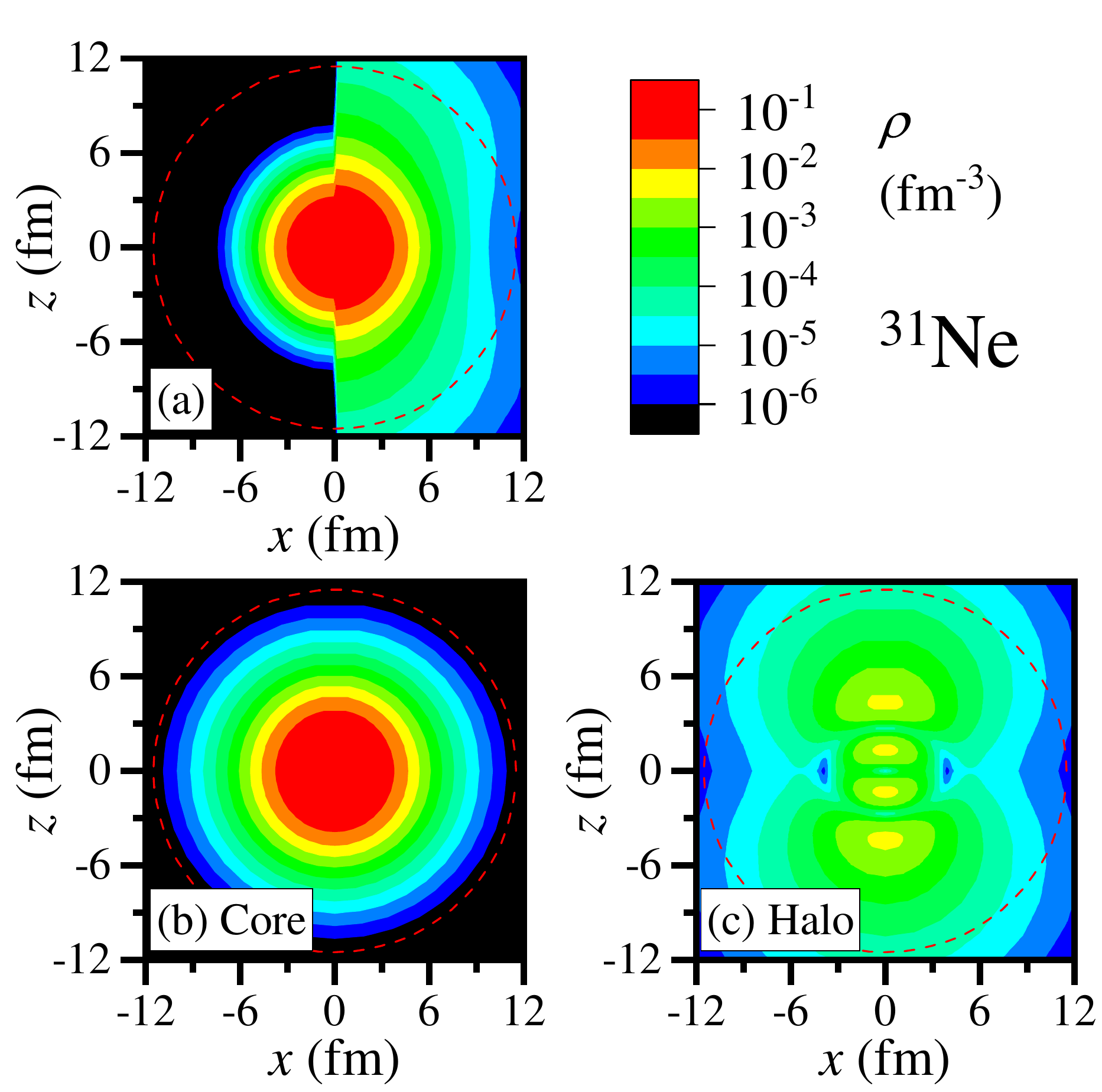}
  \caption{ Density distributions on the $xz$ plane with $y=0$ in $^{31}$Ne for (a) the proton (for $x < 0$) and the neutron (for $x > 0$), and the contributions from the neutron (b) core and (c) halo.
  In each plot, a dotted circle is drawn to guide the eye.
   }
  \label{fig_Ne31dens2D}
\end{figure}

Figure \ref{fig_Ne31dens2D}(a) shows the proton and neutron density distributions for $^{31}$Ne.
Due to the large neutron excess and neutron halo, the distribution of neutron density is much farther than that of proton.
The decomposed neutron densities for the core and the halo are shown respectively in Figs.~\ref{fig_Ne31dens2D}(b) and (c).
In Fig.~\ref{fig_Ne31dens2D}(b), the density distribution of the core is near-spherical, and concentrated near $r=0$.
Its rms radius $R_{\mathrm{core}}=3.47$ fm, which is remarkably lower than the neutron rms radius $R_n=3.73$ fm of $^{31}$Ne.
The quadrupole deformation of core $\beta_{2,\mathrm{core}}=0.03$, which is different with $\beta_2=0$ of $^{30}$Ne, showing the polarization effect of the odd neutron on the core.
In Fig.~\ref{fig_Ne31dens2D}(c), the density distribution of the halo is significantly prolate and diffuse, which is still on the order of $10^{-5}~\text{fm}^{-3}$ at $r=12$ fm along the symmetry axis.
Its rms radius $R_{\mathrm{halo}}=7.17$ fm and the quadrupole deformation $\beta_{2,\mathrm{halo}}=4.21$ are both much larger than those of the core, showing a strong shape decoupling between the prolate halo and the near-spherical core.


\begin{figure}[htbp]
  \centering
  \includegraphics[width=0.4\linewidth]{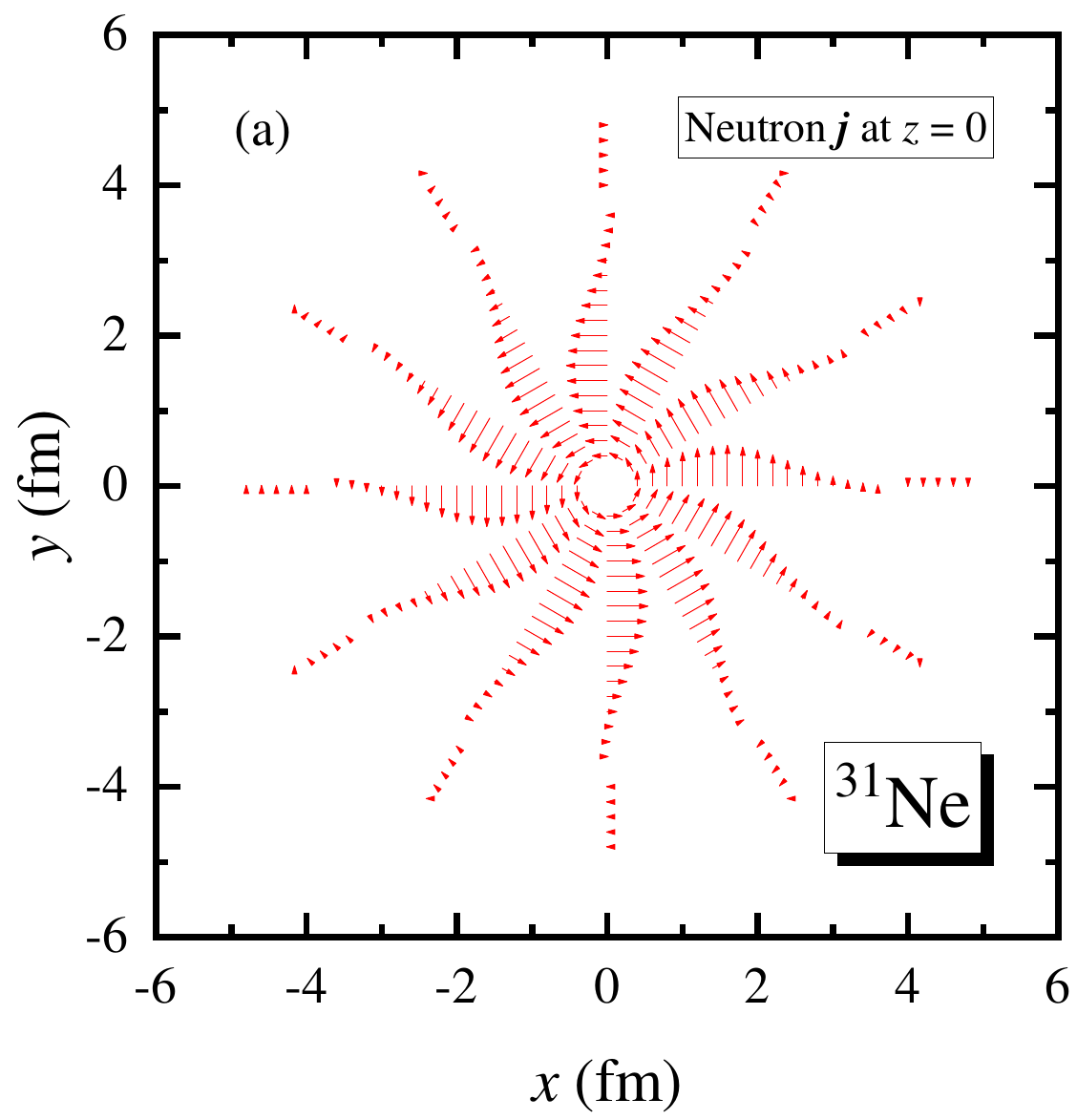}
  \includegraphics[width=0.5\linewidth]{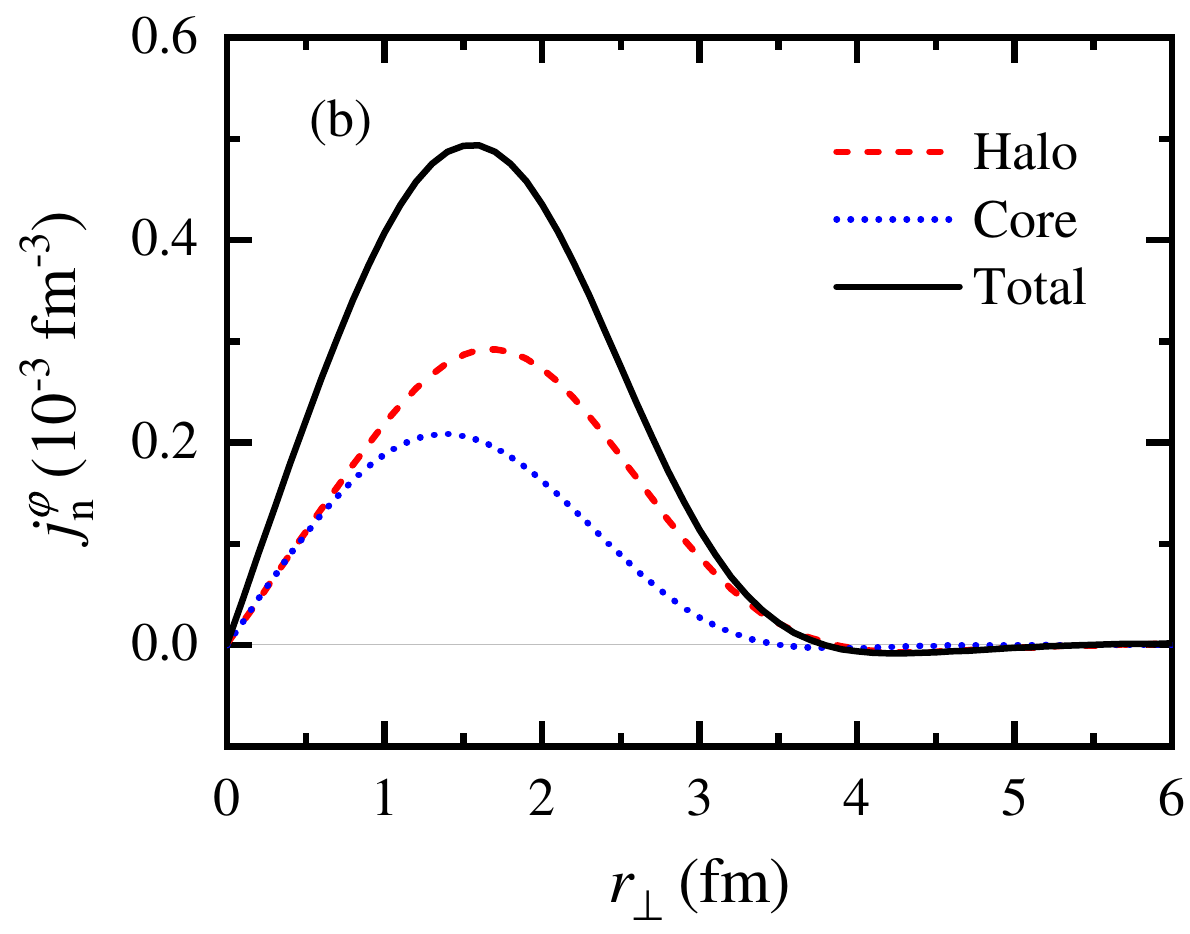}
  \caption{ (a) Distribution of neutron current in the $xy$ plane with $z=0$ in $^{31}$Ne.
  Direction and length of the arrows, respectively, represent orientation and magnitude of current.
  (b) Neutron current as a function of $r_\perp=\sqrt{x^2+y^2}$ with $z=0$ in $^{31}$Ne, as well as its decomposition into the parts contributed from the neutron core and halo.
  }
  \label{fig_Ne31j1Dxy}
\end{figure}

With the self-consistent treatment of nuclear magnetism, the time-odd current is also incorporated in the TODRHBc theory. 
Due to the axial symmetry, the current circles around the symmetry axis and has only the azimuthal component, i.e., $\bm{j} = j^\varphi \bm{\mathrm{e}}_\varphi$ \cite{Hofmann1988PLB}.
Figure \ref{fig_Ne31j1Dxy}(a) shows the distribution of the neutron current in the $xy$ plane with $z=0$ for $^{31}$Ne, with the direction and length of the arrows representing the orientation and magnitude of current, respectively.
There is no current on the symmetry axis at $x=y=0$, and by leaving from the symmetry axis, the vector length increases first, and then decreases.
For a more detailed comparison, Fig.~\ref{fig_Ne31j1Dxy}(b) shows the neutron current as a function of $r_\perp = \sqrt{x^2+y^2}$ with $z=0$, as well as its decomposition into the contributions from the core and halo.
At $r<1$ fm and $r\approx 4$ fm, the contributions from the core and halo are close;
at $1.5 < r <3.5$ fm, the halo part is significantly higher than that of the core;
at $r>4$ fm, $j^{\varphi}$ becomes negative, and the contribution from the halo is still larger than that of the core, which is qualitatively similar to the dominant contribution of halo to neutron density in Fig.~\ref{fig_Ne31dens1D}.
It is also noted that in Fig.~\ref{fig_Ne31dens1D} the density of halo becomes higher than that of core at $r \approx 7.2$ fm, which is much larger than the $r \approx 1$ fm here for neutron current, highlighting the contribution of the halo to the neutron current.


\begin{figure}[htbp]
  \centering
  \includegraphics[width=0.6\linewidth]{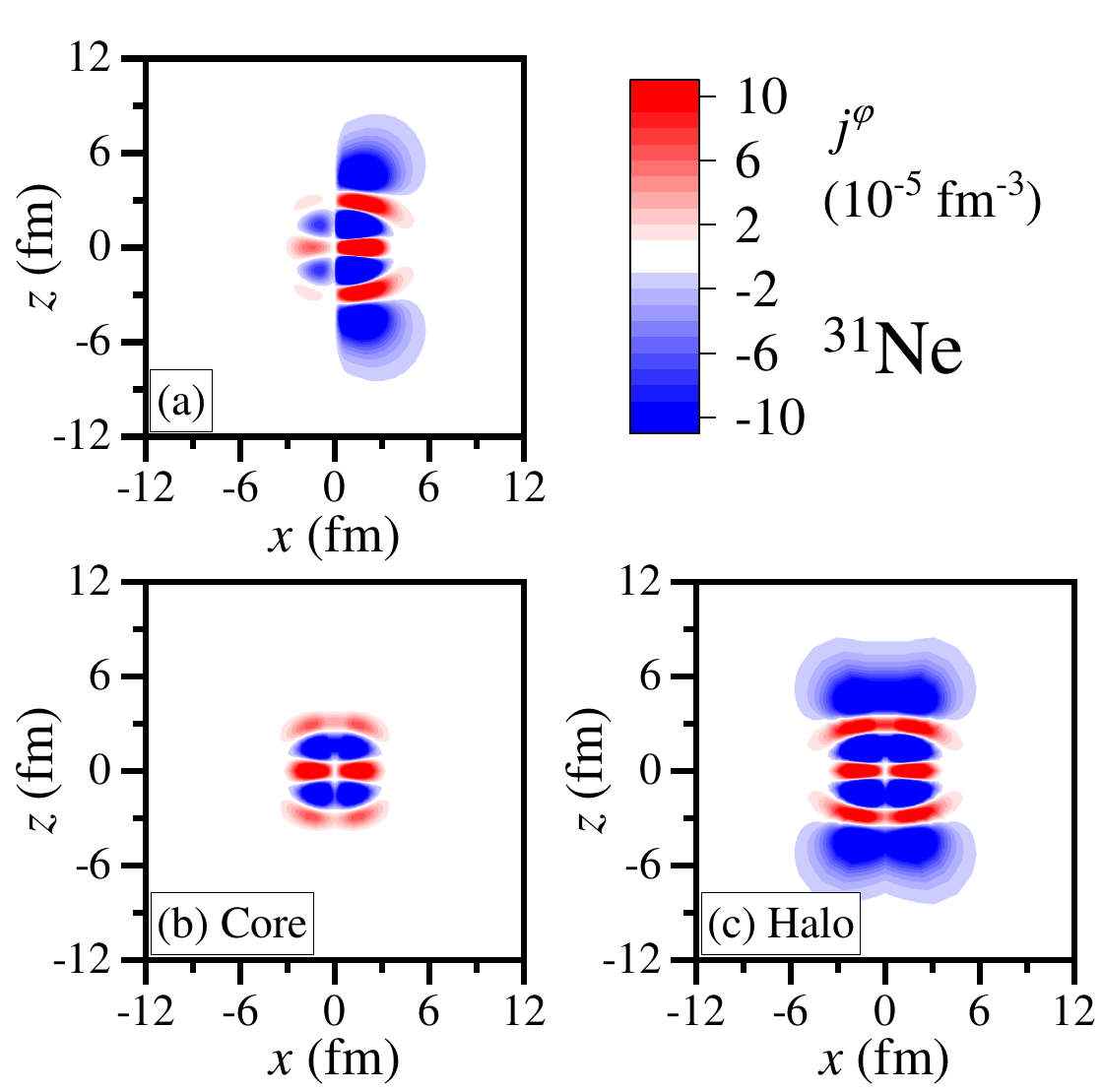}
  \caption{ Distributions of neutron current on the $xz$ plane with $y=0$ in $^{31}$Ne for (a) the proton (for $x < 0$) and the neutron (for $x > 0$), and the contributions from the neutron (b) core and (c) halo.
  }
  \label{fig_Ne31j2D}
\end{figure}

To further investigate the spatial distributions of nucleon currents, Fig.~\ref{fig_Ne31j2D}(a) shows the distributions of proton and neutron currents on the $xz$ plane with $y=0$ for $^{31}$Ne.
Considering the axial symmetry, the nucleon current $\bm{j}$ has only the $j^\varphi$ component, and the direction is always perpendicular to the $xz$ plane. 
Due to the unpaired odd neutron, the distribution of neutron current is much farther than that of proton current.
Similar to the density profiles in Fig.~\ref{fig_Ne31dens2D}(a), the distribution of $j^\varphi$ also forms a prolate shape.
Figures \ref{fig_Ne31j2D}(b) and (c) show the contributions to $j^\varphi$ from core and halo, respectively.
The halo component is more diffuse than the core component at large $r$, which is similar to the density in Fig.~\ref{fig_Ne31dens2D}.
However, it is also noted that even near the center of nucleus, the contribution from halo to neutron current is still not smaller than that from core, which is different with the case of density.
Therefore, we can conclude that at most regions in $^{31}$Ne, the halo provides a dominant contribution to the neutron current.


\begin{figure}[htbp]
  \centering
  \includegraphics[width=0.7\linewidth]{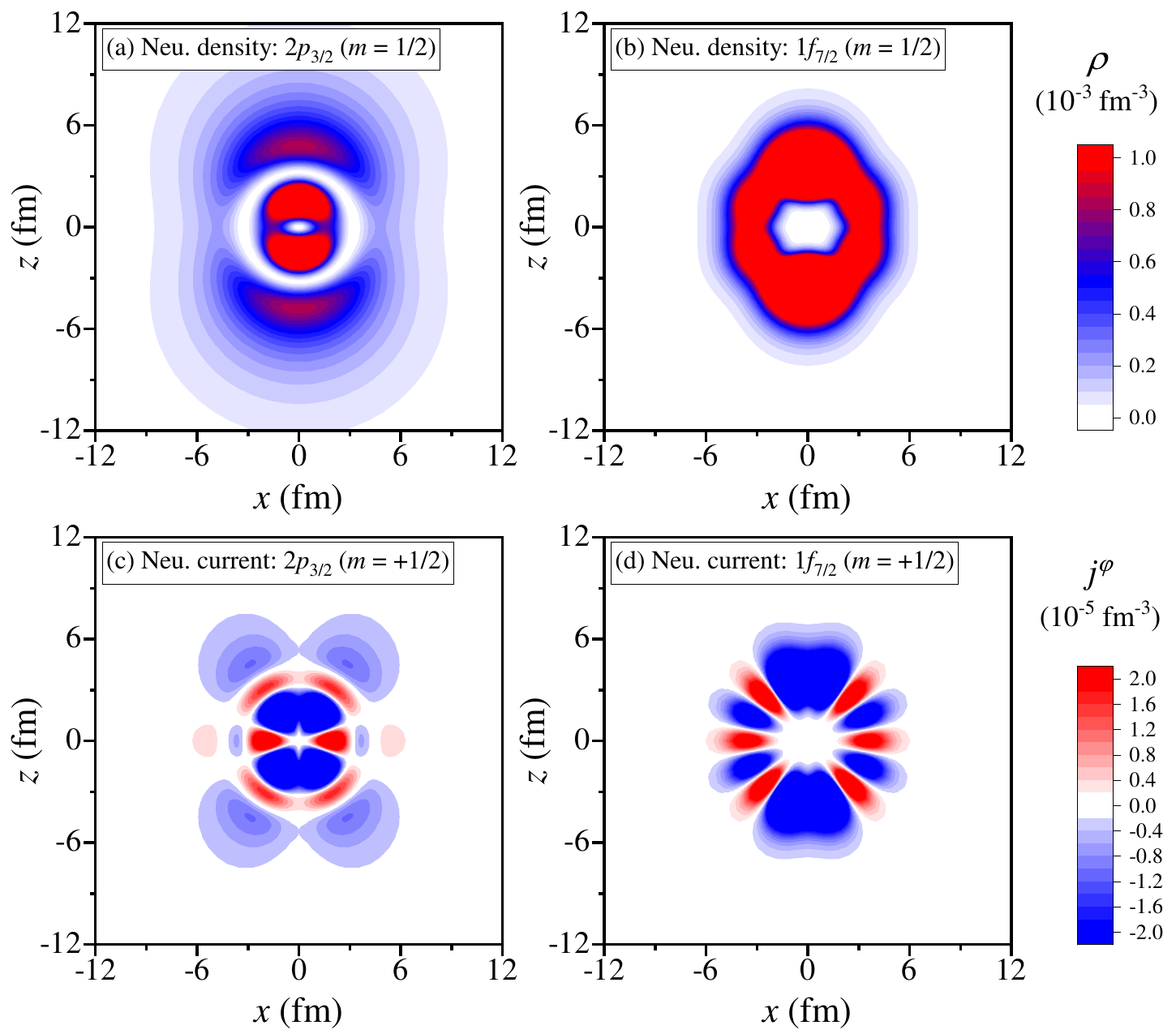}
  \caption{ Distributions of neutron densities on the $xz$ plane with $y=0$ for (a) $2p_{3/2}$ and (b) $1f_{7/2}$ wavefunctions with $m=1/2$ in the DWS basis, and distributions of neutron currents on the $xz$ plane with $y=0$ for (c) $2p_{3/2}$ and (b) $1f_{7/2}$ wavefunctions with $m=+1/2$ in the DWS basis.
  }
  \label{fig_DWSden}
\end{figure}

In Fig.~\ref{fig_Ne31j2D} it is noticed that the distribution of $j^\varphi$ can be divided into several layers, and in different layers, the sign of $j^\varphi$, i.e., the direction of current, is different.
Taking the distribution of halo current in Fig.~\ref{fig_Ne31j2D}(c) as an example, $j^\varphi$ is positive near $z=0$, and by increasing $z$, $j^\varphi$ becomes negative at $z\approx 1$ fm and positive at $z\approx 3$ fm, and finally becomes negative again after crossing $z\approx 4$ fm.
In order to understand such layered structure, here we start from the expression of nucleon current $j^\varphi$ for an axially deformed nucleus,
\begin{equation}
	\label{e:jphi}
	j^\varphi = \sum_m \sum_{n\kappa} \sum_{n'\kappa'} \rho_{n\kappa m,n'\kappa'm'} \frac{1}{r^2}
	\left[ G_{n\kappa} F_{n'\kappa'} (\mathrm{i} Y_{jm}^{l\dagger} \sigma_\varphi Y_{j'm'}^{\tilde{l}'}) - F_{n\kappa} G_{n'\kappa'} (\mathrm{i} Y_{jm}^{\tilde{l}\dagger} \sigma_\varphi Y_{j'm'}^{l'}) \right] ,
\end{equation}
where $(n,j,l,m)$ are the radial, total angular momentum, orbital and third-component of angular momentum quantum numbers for DWS basis, respectively, $\kappa=\pi(-1)^{j+1/2}(j+1/2)$ is the relativistic quantum number, $\rho_{n\kappa m,n'\kappa'm}^k$ is the density matrix in the DWS basis, and $(G_{n\kappa},F_{n\kappa})$ are the DWS radial wavefunctions.
Considering in Fig.~\ref{fig_Ne31spl} the main DWS components of the neutron halo orbital are $2p_{3/2}$ and $1f_{7/2}$ with $m=+1/2$, the corresponding distributions of densities and currents are extracted and shown in Fig.~\ref{fig_DWSden}.
In the density distribution of $2p_{3/2}$ in Fig.~\ref{fig_DWSden}(a), the density is zero at $r=0$ and $r\approx 4$ fm, corresponding to the nodes in the radial wavefunctions $G$ and $F$ of $2p_{3/2}$, while for $1f_{7/2}$ in Fig.~\ref{fig_DWSden}(b) the radial node only appears at $r=0$.
For the neutron current $j^\varphi$ of $2p_{3/2}$ in Fig.~\ref{fig_DWSden}(c), with the increase of $r$, the sign transitions near $r\approx 4$ fm also correspond to the nodes in $G$ and $F$ for Eq.~\eqref{e:jphi}.
By increasing $r$ there is no sign transition for $1f_{7/2}$ in Fig.~\ref{fig_DWSden}(d) because its radial wavefunctions $G$ and $F$ have no node except for that at $r=0$.
The angular dependence of $j^\varphi$ is determined by the coupling terms between spinor spherical harmonics and Pauli matrix $\sigma_\varphi$ in Eq.~\eqref{e:jphi}, contributing to the angular distribution in Fig.~\ref{fig_DWSden}.
Combining the contributions from $2p_{3/2}$ and $1f_{7/2}$, the layered structure for the radial and angular distributions of $j^\varphi$ in Fig.~\ref{fig_Ne31j2D}(c) is formed.


\section{Summary}
\label{sec:summary}

In summary, the time-odd deformed relativistic Hartree-Bogoliubov theory in continuum is developed.
The suggested halo nucleus $^{31}$Ne is investigated in the TODRHBc theory with density functional PC-PK1 and the effects of nuclear magnetism are explored.
It is found that the nuclear magnetism brings about extra 0.09 MeV to the binding energy and the splitting of 0 to 0.2 MeV in the canonical single-neutron and single-proton spectra due to the breaking of Kramers degeneracy.
The blocked neutron level with a dominant contribution from $p_{3/2}$ is obtained for $^{31}$Ne, which is in agreement with the experimental spin-parity $J^\pi=3/2^-$.
The blocked level is a weakly-bound one with single-particle energy $\epsilon = -0.087$ MeV, while it is unbound with $\epsilon = 0.049$ MeV if the nuclear magnetism is neglected.
According to the canonical single-particle spectra and density distribution, a prolate one-neutron halo is formed around the near-spherical core in $^{31}$Ne.
In the distributions of vector current in the $xy$ and $xz$ planes, the current circles around the symmetry axis, and in most regions the contribution from the halo is significantly larger than that from the core.
Even near the center of the nucleus ($r<1$ fm), the contributions to the neutron current from the halo and core are close, which is different with the case of neutron density.
A layered structure in the distribution of neutron current is noticed, and studied by extracting the radial and angular behaviors of the main components in the DWS basis.

It has been expected that the nuclear magnetism may transform an unbound proton-rich nucleus in time-even calculations into a bound one \cite{Afanasjev2010PRCnonrot}.
In our work, it is found that on the neutron-rich side, $^{31}$Ne becomes unstable against one-neutron emission according to the single-neutron spectrum when the nuclear magnetism is neglected, while it is bound after including the nuclear magnetism.
Despite the possible dependence on numerical conditions, our results demonstrate a mechanism of nuclear magnetism in exotic nuclei that makes an orbital from unbound to bound and a nucleus more stable.
It is also interesting to further investigate the nuclear magnetism in more exotic nuclei, e.g., the influence on the shape of nuclear chart, the magnetic moment in exotic nuclei, etc., and such works are in progress.

\begin{acknowledgments}

The authors would like to express gratitude to J. Meng for constructive guidance and valuable suggestions. 
Helpful discussions with members of the DRHBc Mass Table Collaboration are highly appreciated.
This work was partly supported by the National Natural Science Foundation of China (Grants No. 11935003, No. 12141501, and No. 12305125), the Natural Science Foundation of Sichuan Province (Grant No. 24NSFSC5910), the State Key Laboratory of Nuclear Physics and Technology, Peking University (Grant No. NPT2023ZX03), and the High-performance Computing Platform of Peking University. 

\end{acknowledgments}

\end{document}